\documentclass[%
 reprint,
superscriptaddress,
 amsmath,amssymb,
 aps,
]{revtex4-2}

\usepackage{graphicx}
\usepackage{dcolumn}
\usepackage{bm}
\usepackage{svg}
\usepackage{comment}
\usepackage{xcolor}

\usepackage[normalem]{ulem}
\usepackage{soul}
\setstcolor{blue}

\usepackage{comment}
\begin{document}

\preprint{APS/123-QED}

\title{Confirmation of stimulated Hawking radiation, but not of black hole lasing}

\author{Jeff Steinhauer}
\affiliation{Department of Physics, Technion—Israel Institute of Technology, Technion City, Haifa 32000, Israel}

\date{\today}

\begin{abstract}
Stimulated Hawking radiation in an analogue black hole in a Bose-Einstein condensate was reported seven years ago, and it was claimed that the stimulation was of the black hole lasing variety. The study was based on observation of rapidly-growing negative-energy waves. We find that the Hawking particles are directly observable in the experimental plots, which confirms the stimulated Hawking radiation. We further verify this result with new measurements. Also, the observed Hawking particles provide a sensitive, background-free probe of the underlying mechanism of the stimulation. The experiment inspired the prediction of the Bogoliubov-Cherenkov-Landau (BCL) mechanism of stimulated Hawking radiation. By computing the Bogoliubov coefficient for Hawking radiation, we find that the stimulation was of the BCL type, rather than black-hole lasing. We further confirm the results with numerical simulations of both black hole lasing and BCL stimulation.
\end{abstract}

\maketitle


Within many analogue black holes \cite{Unruh81,Garay00,Barcelo01,MacherParentani09}, superluminal (actually supersonic) particles can resist the pull of the analogue gravity, and travel outward. When these particles strike the horizon, they stimulate Hawking radiation \cite{CorleyJacobson99}, pairs of Hawking and partner particles \cite{Hawking74,Hawking75}. In this way, the quantity of Hawking radiation exceeds the spontaneous emission. This process was first suggested in the context of black hole lasing, in which the partner particles are reflected from an inner horizon deep inside the analogue black hole, and become the superluminal particles, which then stimulate additional Hawking/partner pairs \cite{CorleyJacobson99,Jain07,LeonhardtPhilbin07,FinazziParentani10,CoutantParentani10,MichelParentani13,deNovaCarusotto16,deNova21,Bermudez17,Bermudez21}. This stimulated Hawking radiation was studied in an experiment in a Bose-Einstein condensate, in which the partner and superluminal particles were observed, and it was asserted that black hole lasing was the source of the superluminal particles \cite{Steinhauer14}. However, this type of measurement is subject to a large, growing background, and it is difficult to isolate the Hawking radiation signal \cite{SteinhauerdeNova17}.

The experiment \cite{Steinhauer14} inspired several theoretical studies \cite{TettamantiCarusotto16,WangJacobson17,WangJacobson17b, SteinhauerdeNova17,TettamantiCarusotto21}, some with alternative explanations to black hole lasing, for the observed partner and superluminal particles. One suggested that superluminal particles can be created at the inner horizon, rather than evolving from the partners as in black-hole lasing \cite{WangJacobson17}. This is the Bogoliubov-Cherenkov-Landau (BCL) mechanism of stimulated Hawking radiation. One of the explanations did not involve Hawking radiation at all, although it relied on reflection from a steep harmonic potential not present in the experiment \cite{TettamantiCarusotto21}.

Stimulated Hawking radiation has been reported in a variety of other analogue black and white holes \cite{Weinfurtner11,Rousseaux08,Euve16,Drori19}, but they involved neither black hole lasing nor the BCL mechanism.

A later experiment with parameters different from those of Ref. \cite{Steinhauer14} showed that the stimulated Hawking/partner pairs can be observed directly, rather than studying the partners and superluminal particles \cite{KolobovSteinhauer21}. In light of this, we look back to the old experiment \cite{Steinhauer14}, and find that the Hawking/partner pairs are observable there also. This confirms the claimed observation of stimulated Hawking radiation. It also gives us a background-free observation tool for studying the Hawking radiation. We show that black hole lasing would not produce sufficient particles to explain the rapidly-growing population of Hawking/partner pairs, suggesting that the BCL mechanism dominates instead.

\begin{figure}
    \includegraphics[width=\columnwidth]{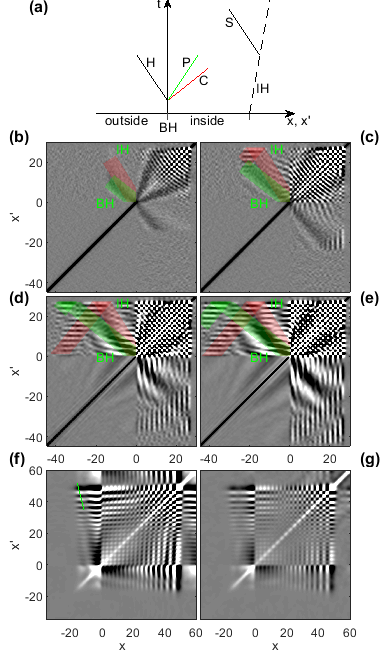}
    \caption{Numerical simulations of the onset of black-hole lasing, and of the experiment. (a) Spacetime diagram of the analogue black hole. The outer and inner horizons are indicated by BH and IH, respectively. H, P, C, and S indicate the Hawking, partner, comoving, and supersonic particles, respectively. (b-e) Numerical simulation of black hole lasing. The correlation function at various times is shown. (b) The green line indicates the spontaneous Hawking/partner correlations emitted from the outer horizon BH. The red line indicates the Hawking/C correlations. (c) The Hawking/partner correlations (green line) are approaching the inner horizon IH. (d) The Hawking/partner correlations (green line) reflect from IH and become Hawking/S correlations with a short wavelength in the vertical direction. (e) The Hawking/S correlations (green line) continue toward BH. (f) Numerical simulation of the experiment, including experimental variations. The green line is a guide to the eye. (g) Numerical simulation of the experiment including quantum fluctuations only.  }
    \label{sim}
\end{figure}

The analogue black hole consists of a Bose-Einstein condensate flowing in the positive $x$-direction. The flow is supersonic for $x>0$, which creates a trapping region for sound quasiparticles, in analogy with the inside of a black hole, as indicated in Fig. \ref{sim}(a). The acceleration at the point $x=0$ is achieved by an applied step potential.

Black hole lasing begins with the spontaneous emission of Hawking radiation from the horizon BH, as illustrated in the spacetime diagram of Fig. \ref{sim}(a). The Hawking particles H carry positive energy, and the partners P carry negative energy. A small quantity of positive energy particles C, which are directed inward in the free-falling frame (comoving with the flowing condensate), are also produced. The P and C particles form a tilted sound cone inside the analogue black hole. Fig. \ref{sim}(b-e) shows a numerical simulation of an analogue black hole, where we combine symmetric outer \cite{Carusotto08} and inner horizons. This is a stationary configuration which suppresses the BCL emission. The correlations between the spontaneous Hawking/partner pairs are seen in Fig. \ref{sim}(b) as a gray band \cite{Balbinot08, Carusotto08}, which is highlighted with a green band. At some point in time, the partners strike the inner horizon of the analogue black hole (IH in Fig. \ref{sim}(a)), and are reflected in the form of supersonic S particles. These particles are analogous to hypothetical superluminal trans-Planckian particles, which travel from the inner horizon outward. This reflection is seen in Fig. \ref{sim}(d), where the green Hawking/partner correlation band reflects from IH, in the form of Hawking/S correlations. The latter correlations have a short wavelength in the vertical direction, since the P mode has become an S mode with large wavenumber. In Fig. \ref{sim}(e), these S particles propagate toward the outer horizon BH. When they will impinge on the outer horizon, they will produce additional Hawking/partner pairs. The process repeats with a period $\tau_{\text{RT}}$, which is the round-trip propagation time between the horizons. The quantity of Hawking radiation thus increases exponentially in time, which is black hole lasing.

For each S particle which impinges on the horizon, $|\beta|^2$ Hawking/partner pairs are produced, where the Bogoliubov coefficient introduced by Hawking \cite{Hawking74} is given by
\begin{equation}
    |\beta|^2=1/(e^{\hbar\omega/k_\text{B}T_\text{H}}-1)
    \label{beta2}
\end{equation}
where $\omega$ is the frequency of the S particle, and $T_\text{H}$ is the Hawking temperature, which is proportional to the (analogue) surface gravity. The number of particles thus increases by a factor $|\beta|^2+1$ every $\tau_{\text{RT}}$. The exponential growth rate for black hole lasing is thus \cite{Steinhauer14}
\begin{equation}
    \tau_\text{L}=\frac{\tau_{\text{RT}}}{\text{ln}(|\beta|^2+1)}
    \label{tauL}
\end{equation}

\begin{figure}
    \includegraphics[width=\columnwidth]{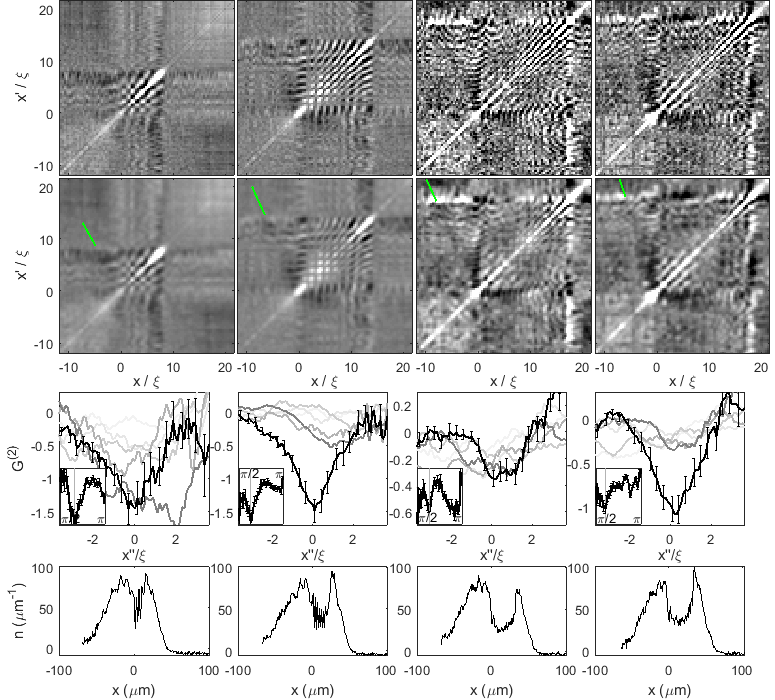}
    \caption{Hawking/partner correlations in the experiment. Correlation functions from Ref. \cite{Steinhauer14} are shown. The first row shows the raw correlation function. The second row is smoothed (convoluted with a Gaussian) to improve the visibility of the Hawking/partner correlation band. The green line indicates the angle of the Hawking-partner correlation band, taken from the inset of the third row. The third row (black curve) shows the profile of the correlation band, integrated along parallel segments at the angle found in the inset.  The gray curves show earlier times, where the darker the gray the later the time.  The inset of the third row shows the angular profile of the smoothed correlation function. The gray line indicates the minimum of the correlation band. The fourth row shows the ensemble-averaged density profile. The first through third columns show small, medium, and large step heights, respectively. The fourth column shows the large step 20 ms later.}
    \label{oldExpt}
\end{figure}

Figure 1(f) shows a more realistic numerical simulation of the experiment. The Hawking/partner correlations due to the stimulated Hawking radiation are visible as a gray band in the direction of the green line. This strong gray band was first discovered in the experimental plots of Ref. \cite{KolobovSteinhauer21}. We can look for this band in the data from Ref. \cite{Steinhauer14}, as seen in Fig. \ref{oldExpt}. The band is clear in columns 2-4, in the direction of the green line. It is even more apparent in the profile of the band shown in the third row (black curve and points), which has a negative peak in all columns. This confirms the observation of stimulated Hawking radiation in Ref. \cite{Steinhauer14}.

\begin{figure}
    \includegraphics[width=\columnwidth]{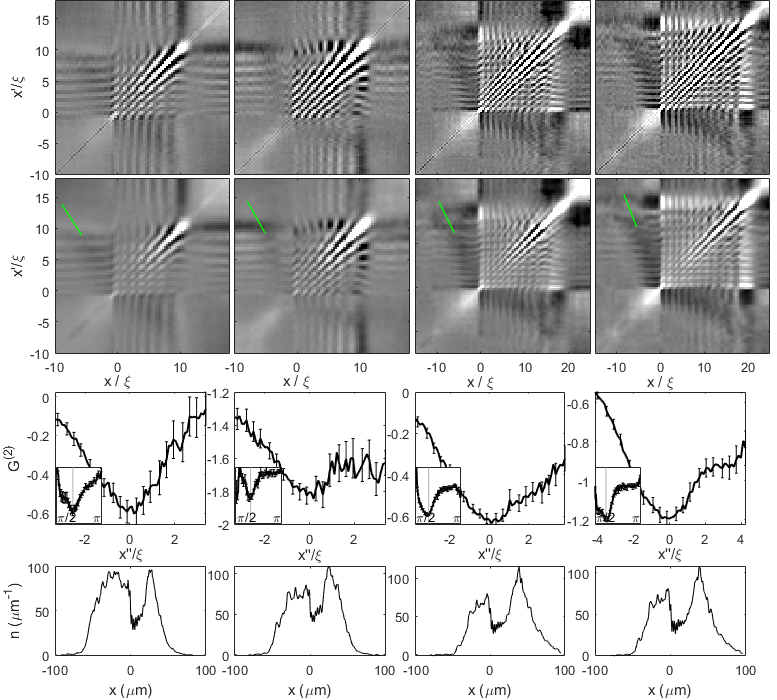}
    \caption{Stimulated Hawking radiation in the new experiment.  The description of the rows is the same as in Fig. \ref{oldExpt}.  From left to right, the 4 columns show step heights of 0.59, 0.59, 0.60, and 0.65, respectively.  The second column shows 66.5 ms later than the first column.}
    \label{newExpt}
\end{figure}

We have the opportunity to verify the stimulated Hawking radiation of Fig. \ref{oldExpt} in a greatly improved experimental apparatus. Fig. \ref{newExpt} shows new data gathered 5 years later, using the experimental apparatus of Ref. \cite{KolobovSteinhauer21}. The experimental parameters are the same as in Ref. \cite{KolobovSteinhauer21}, with the exception of the step height, which is decreased by a factor of approximately 0.6 in order to give similar density profiles (row 4) to Fig. \ref{oldExpt}.  One indeed sees a clear Hawking/partner correlation band in all columns of \ref{newExpt}. However, with the smaller step height, the correlation band is less pronounced than in Ref. \cite{KolobovSteinhauer21}. This trend is also seen in Fig. \ref{oldExpt}, in which the visibility of the correlation band increases from left to right.

\begin{figure}
    \includegraphics[width=\columnwidth]{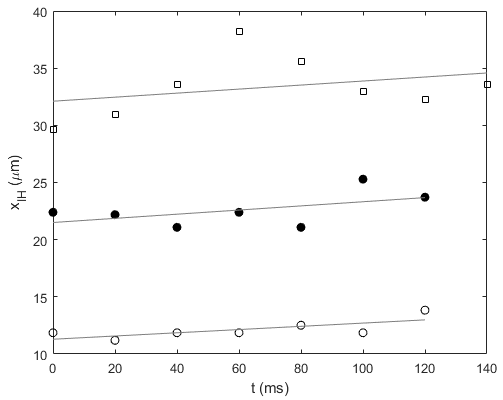}
    \caption{Motion of the inner horizon in Ref. \cite{Steinhauer14}. The position of the inner horizon is shown for the small step (open circles), medium step (filled circles), and large step (squares). The lines are linear fits.}
    \label{horizon}
\end{figure}

We can compare the growth of the Hawking/partner correlation band in Fig. \ref{oldExpt} to the expectation for black hole lasing. Simulations suggest that the inner horizon should move away from the outer horizon with a velocity $v_\text{IH}$, as illustrated in Fig. 1(a) \cite{WangJacobson17}. In black hole lasing, the partners reflect from this moving horizon, which causes a Doppler shift $k_0v_\text{IH}$ in the frequency of the S particles, where $k_0$ is their large wavenumber. This increase in frequency will decrease the rate of black hole lasing by Eq's. \ref{beta2} and \ref{tauL}. We can evaluate $v_\text{IH}$ in the experimental data of Ref. \cite{Steinhauer14}. Fig. \ref{horizon} shows the position of the inner horizon as a function of time. The data shows a slight positive slope. The linear fits give $v_\text{IH}$ = 14(8), 18(17), and 18(26) $\mu$m s$^{-1}$ for the small, medium, and large steps, respectively. The values of $k_0$ are 1.7, 2.3, and 2.8 $\mu$m$^{-1}$. Thus, the incoming frequencies are $\omega/2\pi$ = 4(2), 7(6), and 8(12) Hz. If we assume that $k_\text{B}T_\text{H}$= 0.35 nK as in Ref. \cite{KolobovSteinhauer21}, then $|\beta|^2\approx$ 2, 0.7, and 0.5, by Eq. \ref{beta2}. Combining this with $\tau_\text{RT}$ = 57, 55, and 78 ms from Ref. \cite{Steinhauer14}, the lasing times are $\tau_\text{L}\approx$ 60, 110, and 180 ms, by Eq. \ref{tauL}. These values are 2 orders of magnitude smaller than a similar calculation for Ref. \cite{KolobovSteinhauer21}, due to $v_\text{IH}$ which is 5 times smaller. By careful inspection of the third row of Fig. \ref{oldExpt}, one sees that the Hawking/partner correlations grow significantly in 20 ms. Thus, $\tau_\text{L}$ is too long to explain the growth, and one concludes that the stimulated Hawking radiation is due to the BCL mechanism. In other words, although the lasing is 2 orders of magnitude faster than in Ref. \cite{KolobovSteinhauer21}, it is still too slow to explain the observations.

This conclusion is consistent with previous numerical simulations \cite{WangJacobson17,WangJacobson17b}, which found a lack of black hole lasing in Ref. \cite{Steinhauer14}. However, one of these simulations \cite{WangJacobson17} had a $v_\text{IH}$ which was 5 times faster than in \cite{Steinhauer14}, which would greatly increase $\tau_\text{L}$. The simulation in Ref. \cite{WangJacobson17b} was likely improved in this respect, and it reached the same conclusion.

As explained in Ref. \cite{KolobovSteinhauer21}, we can use a simulation to study the Hawking/partner correlation band, and differentiate between black hole lasing and BCL stimulation, since the former is stochastic, while the latter is deterministic. Fig. \ref{sim}(f) shows our numerical simulation, with parameters similar to Ref. \cite{Steinhauer14}, with the exception of $v_\text{IH}$, which is 3 times faster. The simulation includes run-to-run variations in the total number of particles, which allows a deterministic process to appear in the correlation function \cite{WangJacobson17b}. The correlation band of stimulated Hawking radiation is visible in the direction of the green line. On the other hand, no such correlation band is visible in Fig. \ref{sim}(g), which has quantum fluctuations only. Since the number variations are required to create the Hawking/partner correlation band, the dominant process is the deterministic BCL mechanism.

In conclusion, stimulated Hawking radiation is confirmed in the 2014 experiment, by direct observation of the Hawking/partner pair correlations. The result is further supported by new measurements. The observed pairs provide a sensitive test of the mechanism of stimulated Hawking radiation. We find that the rate of black hole lasing would be too small to explain the pair production, which suggests that the BCL mechanism dominates. This is supported by numerical simulations.

\begin{acknowledgments}
I thank J. R. M. de Nova for performing the numerical simulations. This work was supported by the Israel Science Foundation.
\end{acknowledgments}

\appendix


\bibliography{smallStep}

\end{document}